# Understanding fast macroscale fracture from microcrack post mortem patterns

C. Guerra,[1,2] J. Scheibert,[1,3,4,5] D. Bonamy,[1] and D. Dalmas[3]

1. *CEA, IRAMIS, SPCSI, Group Complex Systems & Fracture, F-91191 Gif sur Yvette, France*

2. *Facultad de Ingeniería Mecánica y Eléctrica, Universidad Autónoma de Nuevo León, Avenida Universidad, S/N, Ciudad Universitaria, C.P. 66450, San Nicolás de los Garza, NL, Mexico*

3. *Unité Mixte CNRS/Saint-Gobain, Surface du Verre et Interfaces, 39 Quai Lucien Lefranc, 93303 Aubervilliers cedex, France*

4. *Physics of Geological Processes, University of Oslo, P.O. Box 1048 Blindern, 0316 Oslo, Norway*

5. *Laboratoire de Tribologie et Dynamique des Systèmes, CNRS, Ecole Centrale de Lyon, 36 Avenue Guy de Collongue, 69134 Ecully, France*

*Dynamic crack propagation drives catastrophic solid failures. In many amorphous brittle materials, sufficiently fast crack growth involves small-scale, high-frequency microcracking damage localized near the crack tip. The ultra-fast dynamics of microcrack nucleation, growth and coalescence is inaccessible experimentally and fast crack propagation was therefore studied only as a macroscale average. Here, we overcome this limitation in polymethylmethacrylate, the archetype of brittle amorphous materials: We reconstruct the complete spatio-temporal microcracking dynamics, with micrometer / nanosecond resolution, through post mortem analysis of the fracture surfaces. We find that all individual microcracks propagate at the same low, load-independent, velocity. Collectively, the main effect of microcracks is not to slow down fracture by increasing the energy required for crack propagation, as commonly believed, but on the contrary to boost the macroscale velocity through an acceleration factor selected on geometric grounds. Our results emphasize the key role of damage-related internal variables in the selection of macroscale fracture dynamics.*

The fracture of brittle amorphous materials is usually described using the linear elastic fracture mechanics (LEFM) framework [1–4], which considers the straight propagation of a single smooth crack. All dissipative processes (*e.g.* plastic deformation or bond breaking) are assumed to be localized in a small zone around the crack tip (fracture process zone, FPZ). Crack velocity, $v$, is then predicted from the balance between the flux of mechanical energy relased from the surrounding elastic material into the FPZ [5], and the dissipation rate within this zone. The former is computable within continuum theory and connects to the stress intensity factor, $K$, which describes the macroscopic forcing applying on the crack tip and depends on the external loading and specimen geometry only. The dissipation rate is quantified by the fracture energy, $\Gamma$, required to expose a new unit area of cracked surfaces, to be measured experimentally. The resulting equation of motion reads [1] $\Gamma \approx \left(1 - v/c_R\right)\frac{K^2}{E}$ where $E$ and $c_R$ denote the material's Young's modulus and Rayleigh wave speed, respectively.

Polymethylmethacrylate (PMMA) is often considered as the archetype of nominally brittle materials and, therefore, has been one of the most widely used materials against which theories have been confronted from the early stages of fracture mechanics. Yet, in PMMA, single smooth cracks are actually observed for slow propagation only. Fast enough cracks ($v > v_a \sim 0.2\ c_R$ [6]) propagate through the nucleation, growth and coalescence, in the fracture plane, of individual microcracks [6–9]. Cracks faster than $v_b \sim 0.4\ c_R$ also involve aborted out-of-plane secondary cracks known as microbranches [2,10,11], which prevent LEFM from being applicable [11]. LEFM has been shown to agree with experiments as long as no microbranch is involved [6,11–13], i.e. even in the presence of microcracks, provided a suitable velocity dependence of the fracture energy, $\Gamma(v)$, is prescribed [6,11].

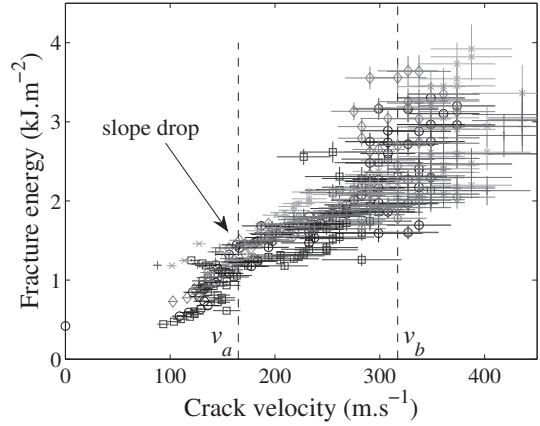

FIG. 1. Fracture energy $\Gamma$ as a function of macroscale crack velocity $v$ (adapted from [6]). Different symbols correspond to different experiments. The two vertical dashed lines correspond to $v_a$ (microcracking onset) and $v_b$ (microbranching onset). Below $v_b$, all the experimental points collapse onto a a single $\Gamma(v)$ curve. The slope of this curve exhibits a drop at $v_a$.

Recent experiments using PMMA showed that, above $v_a$, the slope of $\Gamma(v)$ drops [6] (see Fig. 1), suggesting that microcracks make macroscale cracks dissipate less or/and propagate faster than a single crack would. This is at odds with the common view that damage through opening mode microcracks slows down crack propagation by increasing energy dissipation [3,7,14]. Understanding this counter-intuitive behaviour requires unravelling the coupling between (i) the space-time dynamics of damage at the FPZ scale and (ii) the crack dynamics at the macroscale. The time interval between two successive microcrack nucleation events is typically a few tens of nanoseconds. This makes real-time local measurements of microcracking dynamics beyond current researchers' reach. Hence, fast crack propagation has been studied only through measurements of the average dynamics of the macroscopic crack front [5–8,10–16].

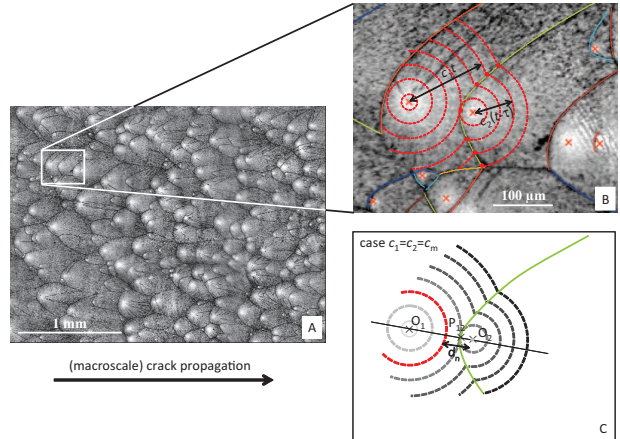

FIG. 2. Fractographic signature of microcracking in the dynamic fracture of PMMA. (A) Typical fractographic microscope image ($K \cong 3.10 \pm 0.05$ MPa.m$^{1/2}$). Bright regions correspond to microcrack nucleation centers (Materials and Methods). (B) Red dashed circle arcs sketch successive front locations of two interacting microcracks (nucleated at $t=0$ and $t=\tau$) growing radially at speeds $c_1$ and $c_2$. Fitting of fractographic branches (color lines) with a geometrical model (Eq. (S1)) allows measuring $c_2/c_1$. (C) When $c_2 = c_1$, markings (green line) are conic branches (Eq. (S2)) and the distance $d_n$ between the triggered microcrack center and the triggering front at the nucleation time $t = \tau$ (highlighted in red) is twice the apex-to-focus distance, $O_2P_{12}$.

Quantitative fractography is an appealing tool to probe microscale damage mechanisms. Fracture surfaces are indeed known to record fracture processes down to the nanoscale [9,17]. In particular, in many materials including PMMA (Fig. 2A), microcracks leave characteristic conic-like markings on fracture surfaces [3,6–9]. These patterns are commonly understood through a geometrical model first developed in [18] and improved in *e.g.* [7,19]. In this model, each conic-like marking corresponds to the intersection of two penny-shaped microcracks, nucleated at point-like nucleation centers and growing at speeds $c_1$ and $c_2$ along two slightly different planes (Fig. 2B). The numerical implementation of this model demonstrated that microcracking is responsible for some of the complexity of macroscopic crack growth [7], e.g. mist fracture surfaces decorated by conic-like markings and strong fluctuations in the velocity signal, $v(t)$. However, the agreement remained only qualitative because simplifying prescriptions were used for the microcracking dynamics [7], namely (i) the location of nucleation centers, (ii) $c_2/c_1$ and (iii) the nucleation criterion.

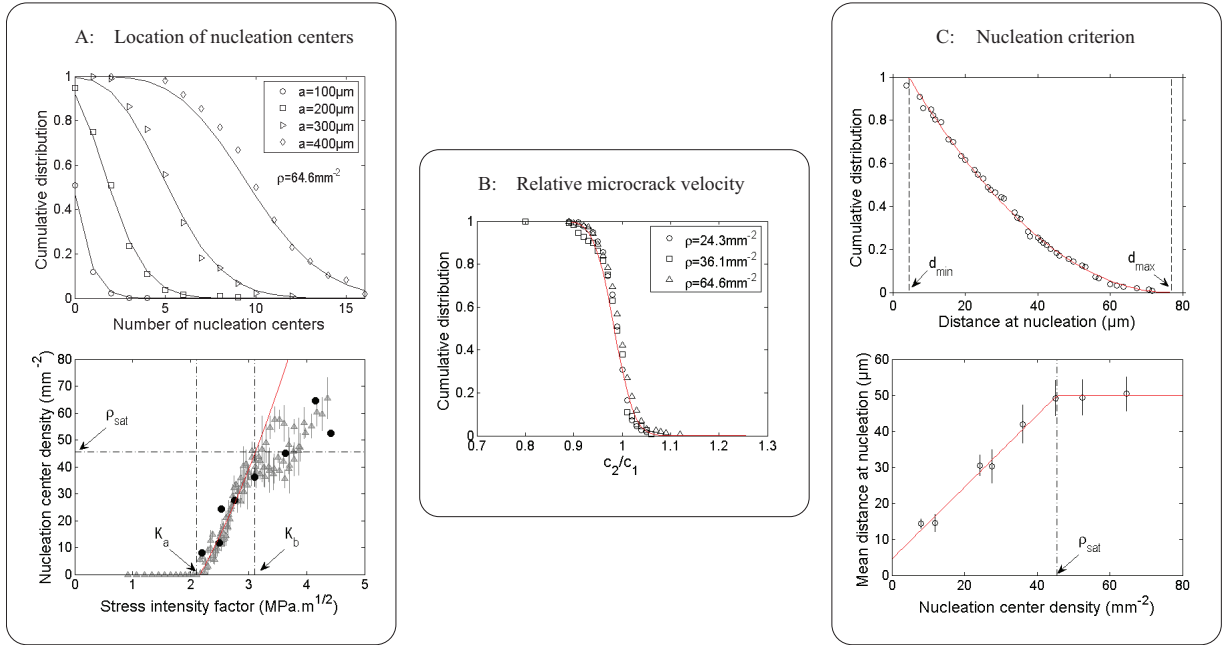

FIG. 3. Microcrack dynamics deduced from fracture surfaces. (A) Top: cumulative distributions for the number of nucleation centers in square regions of size $a$, for $K \cong 4.15 \pm 0.07$ MPa.m$^{1/2}$. Solid lines: Poisson function with parameter $\rho a^2$. The fitting parameter $\rho$ is $a$-independent, indicating homogeneous uncorrelated random distribution with mean surface density $\rho$. Similar results hold for all $K$ (Fig. S3). Bottom: $\rho(K)$ curve (black disks) superimposed to that obtained from the data reported in [6] (gray triangles, vertical lines indicate standard deviation). Red line: fit using Eq. (S3) in the range $K_a=2.1$ MPa.m$^{1/2} < K < K_b=3.1$ MPa.m$^{1/2}$ (see *SI text*). $\rho_{sat} \cong 45.5$ mm$^{-2}$. (B) Cumulative distribution, for various $\rho$, of the velocity ratio $c_2/c_1$ obtained via fitting fractographic branches using the geometrical model (see Eq. (S1) and *SI text*). Red line: Fitted normal distribution (average 0.98 and standard deviation 0.03). Similar results hold for each $\rho$ (Fig. S4). (C) Top: cumulative distribution of $d_n$ for $\rho = 27.5$ mm$^{-2}$. Red line: best two-parameters fit $P(d_n) = ((d_{max} - d_n)/(d_{max} - d_{min}))2$. Here $d_{min}=4$ μm, $d_{max}=77$ μm. Similar fits hold for all $\rho$ (Fig. S5). Bottom: mean distance at nucleation $\overline{d_n}$ as a function of $\rho$. Error bars: $\pm$ one standard deviation. Red line: fit using Eq. (S4) up to a saturating value $\overline{d_n} \cong 50$ μm reached at $\rho_{sat}$ (see *SI text*).

EXPERIMENTAL DETERMINATION OF THE DYNAMICS OF INDIVIDUAL MICROCRACKS

Here, we determine experimentally the microscopic rules for the nucleation and growth of microcracks, by analyzing the morphology of each individual conic-like marking on different millimeter-sized fracture surfaces (see e. g. Fig. 2A) corresponding to different $K$ (*i.e.* to different $v$ in the range 0.23-0.49 $c_R$) (Materials and Methods). We first find that, irrespective of $K$, the spatial distribution of nucleation centers is Poissonian (see Fig. 3A, top and Fig. S3), *i.e.* the centers are homogeneously and randomly distributed in space, without correlation. This is consistent with the usual view that microcracks nucleate at some preexisting weak defects randomly distributed within the material's volume, when a crack tip running in their vicinity sufficiently enhances the stress field [3,9]. The increase in mean surface density of nucleation centers, $\rho$, with $K$ (Fig. 3A, bottom) is attributed to the increase in FPZ size with $K$, which yields more volume defects turning into microcracks (see [6] and SI text). Because $\rho$ completely characterizes Poisson distributions, it will be used hereafter as the parameter as a function of which the various quantities will be plotted.

Stationarity of macroscopic crack propagation at the scale of each millimetric-sized image requires the ratio $c_2/c_1$ of the velocities of two successive microcracks to be 1, on average. A smaller (larger) value would indeed produce a decelerating (accelerating) macroscale crack. This requirement has consequences on the geometry of conic-like markings (see SI text and Fig. S2), which were checked: We fitted all individual markings with the shape predicted using the geometrical model (see Fig. 2B and SI text), with $c_2/c_1$ being the only adjustable parameter. Irrespective of $\rho$, $c_2/c_1$ is found equal to 1 within 4% standard deviation (see Fig. 3B and Fig. S4). In the following, we will neglect the slight dispersion of $c_2/c_1$ and consider that, for any given $\rho$, all microcracks propagate at the same velocity: $c_2 = c_1 = c_m$, where $c_m$ denotes the speed of individual microcracks and a priori depends on the macroscopic external loading $K$ (or equivalently on $\rho$).

In these conditions, the intersection between two microcracks is a true conic. Its focus coincides with the nucleation center of the triggered microcrack, and the apex-to-focus distance is half the distance $d_n$ between the triggering front and the triggered center at the instant of nucleation (see Fig. 2C and SI text). Hence, $d_n$ defines the nucleation criterion. Its cumulative distribution is well fitted by a two-parameters parabolic function, irrespective of $\rho$ (Fig. 3C, top and Fig. S5). Variations of the mean value $\overline{d_n}$ with $\rho$ exhibit two regimes: an initial linear increase followed by a saturating plateau, when $\rho$ exceeds a value $\rho_{\text{sat}}$ (Fig. 3C, bottom). The linear behaviour comes from the fact that $\overline{d_n}$ and $\rho$ both scale linearly with the FPZ size (see SI text). The transition is understood as the point where $\overline{d_n}$ becomes comparable with the mean distance between nucleation centers (see SI text and Fig. S6).

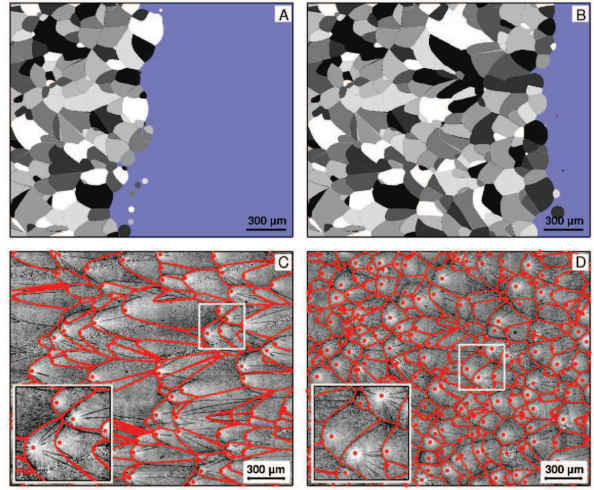

FIG. 4. Deterministic reconstruction of microscale damage and fracture processes. (A-B): successive snapshots of the reconstructed crack propagation and associated conic markings for $\rho$ =64.6 mm$^{-2}$. Crack propagates from left to right. (C-D): Fracture surface images (grey level) for (C) $\rho$ =27.5 mm$^{-2}$ ($K \cong 2.77$ MPa.m$^{1/2}$) and (D) $\rho$ =64.6 mm$^{-2}$ ($K \cong 4.18$ MPa.m$^{1/2}$) compared to the reconstructed conic markings (red lines). Red dots indicate nucleation centers.

## DETERMINISTIC RECONSTRUCTION OF THE MICROCRACKING DAMAGE HISTORY

The analyses performed up to now permit a full characterization of the statistics of microcrack nucleation, growth and coalescence within the FPZ. To unravel how this FPZ quantitatively operates to relate the macroscale crack velocity, $v$, to the microscale velocity, $c_m$, of individual microcracks, we feed the geometrical model with the observed locations of all individual nucleation centers and the corresponding distances at nucleation, $d_n$. We then simulate the space-time evolution of the fracturing process with the constraint that all microcracks propagate at the same velocity $c_1 = c_2 = c_m$ (Materials and Methods). Note that, at this point, $c_m$ is constant within the FPZ but can depend on $K$ (or equivalently on $\rho$). Figures 4A-B show typical snapshots of the simulated crack dynamics (see supporting movies). Apart from edge effects (see SI text and Fig. S7), the matching between the experimental conics and the simulated ones is quite satisfactory (Fig. 4C-D) for all values of $\rho$. As expected, the simulated dynamics thus provide a deterministic reconstruction of the ultra-fast microcracking dynamics. The spatial resolution of $\sim 2$ μm (pixel size) and the time resolution of $\sim 10$ ns (pixel size divided by $c_m$, demonstrated hereafter to be a load-independent constant close to a value $c_m \sim 200$ m/s) are much beyond standard experimental mechanics methods like acoustic emission or fast imaging. Similar deterministic nucleation and geometrical growth models are used in a broad range of fields including metallurgy [20], biology [21] and superconductivity [22].

MACROSCALE CRACK DYNAMICS

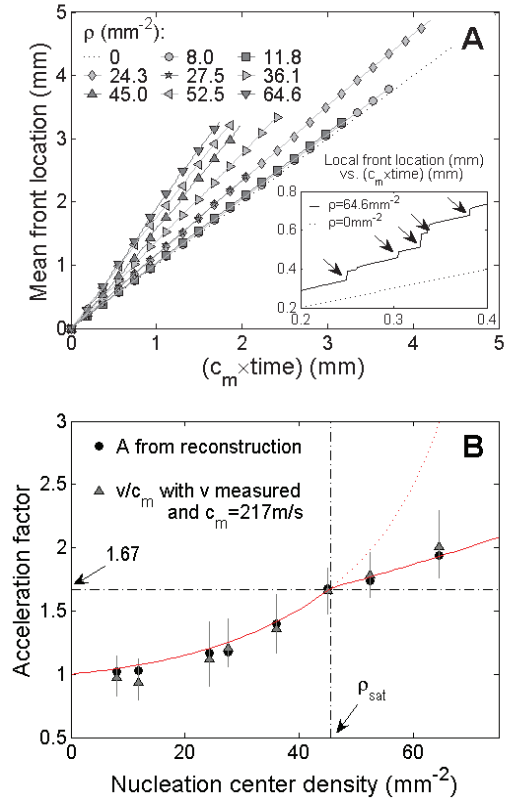

FIG. 5. From slow microcracks to fast collective macroscopic crack motion. (A) Time evolution (scaled by $c_m$) of the average location of the simulated crack front for various $\rho$. The fitted slopes of these curves define the acceleration factor $A$. $A=1$ for $\rho=0$. Inset: Evolution of the location of a single point of the simulated front, for $\rho=64.6$mm$^2$, together with that expected for $\rho=0$ (slope 1). Jumps correspond to coalescence events with microcracks. Between jumps, the slope is close to 1. (B) Black dots: Evolution of the reconstructed acceleration factor $A$ as a function of $\rho$. Triangles: Ratio of the measured macroscopic crack speed, $v$, over the microscopic velocity fitted to be $c_m = 217 \pm 3$ m/s. Error bars indicate the minimum and maximum measured velocities within the considered fractographic image. Thick red line: Eq. 1 with $b=1.19 \pm 0.02$. A change in regime occurs for $\rho = \rho_{sat}$ (vertical dashed line), which corresponds to a velocity of 1.67 $c_m$ (horizontal dashed line). $\pm$ stands for 95% confidence interval.

To shed light on the macroscale effect of microcracking damage, we now focus on the time evolution of the average location of the simulated crack front (Fig. 5A). For each $\rho$, this evolution is linear, meaning that the average front has a constant velocity, $A \times c_m$, the value of which was found insensitive to edge effects (see SI text and Fig. S7). Figure 5B shows that the acceleration factor $A$ equals 1 only for $\rho=0$, and then increases with $\rho$. The time evolution of the position of a single point of the front (Fig. 5A) sheds light on the origin of this effective acceleration. The point motion is jerky, with sudden jumps corresponding to microcrack coalescence events, and the velocity between jumps is close to $c_m$. Hence, as the rate of coalescence events increases with $\rho$, $A$ also increases with $\rho$. A simple mean-field lattice model, which evaluates the rate of coalescence events, yields (see SI text and Fig. S8):

$$A = \frac{1}{1 - b\,\overline{d_n}\,\sqrt{\rho}} \qquad (1)$$

where $b$ is a numerical factor $\cong 1$. This equation, when combined with the observed evolution of $\overline{d_n}$ with $\rho$ (Fig. 3C, bottom, red line), gives the red line in Fig. 5B, which is in very good agreement with reconstructed velocities.

The question remains of the possible dependence of $c_m$ with $\rho$. Figure 5B shows that, if we chose a $\rho$-independent $c_m = 217 \pm 3$ m/s = $0.24 \pm 0.01\ c_R$, the $\rho$-dependence of the reconstructed acceleration factor $A$ is identical to the ratio of the measured macroscopic crack speed, $v$, over $c_m$. This means that the propagation speed of microcracks is not only identical for two successive microcracks, but also all along the crack path, irrespective of $\rho$ - and hence of $K$. Note that $c_m$ is found very close to the maximum speed, $\cong 204$ m/s or $\cong 0.23\ c_R$, of individual crack fronts in PMMA originating from the fracture energy increase with FPZ size [6]. The change in the $\rho$-dependence of $A$ observed at $\rho_{sat}$ in Fig. 5B corresponds to a macroscopic crack speed $\cong 1.67\ c_m \cong 0.41\ c_R$. This velocity is very close to the onset of the microbranching instability (0.36 $c_R$ in PMMA [11]), which suggests that this instability could be related to the steric effect responsible for the saturation of $\overline{d_n}(\rho)$ above $\rho_{sat}$.

DISCUSSION

In dynamic fracture, the relationship between the opening force and the speed at which a macroscopic crack moves forward is controlled by dissipative and nonlinear processes that develop at the microscale within the FPZ. The space and time scales associated with the FPZ dynamics are usually too small to enable a real time and space monitoring of these processes. Here, we demonstrate that such a detailed monitoring is actually possible in PMMA, the archetype of nominally brittle materials, by analysing post mortem the patterns left on fracture surfaces by microcracking damage.

Our results show that, in PMMA, the true local propagation speed of single cracks is limited to a fairly low value $c_m$, about $0.23 c_R$, while the apparent speed, $v$, measured at the continuum-level scale can be much higher. When $v \geq c_m$, the macroscopic crack is actually found to progress through the coalescence of microcracks, all growing at the same constant velocity $c_m$. The main effect of microcracking damage, therefore, is not, as commonly believed [3, 7, 14], to slow down fracture by increasing the energy required to further propagate a crack, but on the contrary to boost the macroscopic (group) crack velocity to a value larger than what would have been obtained in their absence.

We conjecture that the limiting value $c_m$ of the local crack speed is set by the material-dependent dissipative and non-linear processes that develop in the highly stressed/strained zones in the very vicinity of the (micro)crack tips, like e.g. thermal [23], viscoelastic [15, 24] or hyperelastic [5, 25] processes. As for the subsequent boost from $c_m$ to the continuum-level scale velocity $v$, it is shown here to take the form of a purely geometric factor controlled by two microscopic quantities: (i) The density of nucleation centers $\rho$ and (ii) the mean distance at nucleation $d_n$. These two internal variables characterize the damaging state, and evolve with the amount of mechanical energy flowing into the FPZ. As such, they are material-dependent functions of the external loading $K$, the knowledge of which permits to fully relate $v$ and $c_m$.

This enhanced description of dynamic brittle fracture, demonstrated on PMMA, can likely be extended qualitatively to all materials involving propagation-triggered microcracks, e.g. oxyde glass [3, 26], polymeric glasses [3, 9], polycrystals [3], rocks [27, 28] and bones [29]. Further work is required to check this conjecture, and subsequently to quantitatively determine how $c_m$, $\rho$ and $\overline{d_n}$ are selected in these materials. From the geometric nature of the acceleration factor, we also anticipate that fast macroscopic cracks in other fracture modes could similarly originate from the collective motion of many slow microcracks.

MATERIALS AND METHODS

Experiments

Fracture surfaces were obtained from the experiments described in reference [6]. Dynamic cracks were driven in PMMA (Young's modulus $E = 2.8$ GPa and Poisson's ratio $\nu = 0.36$, yielding a Rayleigh wave speed $c_R = 880$ m/s) using the Wedge-Splitting geometry sketched in Fig. S1. Specimens were prepared from parallelepipeds of size $140 \times 125 \times 15$ mm$^3$ in the propagation, loading, and thickness directions, respectively. Subsequently, a notch was formed (i) by cutting a $25 \times 25$ mm$^2$ rectangle from the middle of one of the $125 \times 15$ mm$^2$ edges; and (ii) by subsequently adding a 10 mm groove deeper into the specimen. A circular hole with a radius ranging between 2 and 8 mm was eventually drilled at the tip of the groove. Two steel jaws equipped with rollers were placed on both sides of the rectangular cut-out and a steel wedge of semi-angle 15 degrees was pushed between them at constant velocity 38 m/s up to crack initiation. Crack speed was measured using a modified version of the potential drop technique: A series of 90 parallel conductive lines (2.4 nm-thick Cr layer covered with 23 nm-thick Au layer), 0.5 mm wide with a period of 1 mm (space accuracy 40 μm) were deposited on one of the two $140 \times 125$ mm$^2$ sides of the specimen, connected in parallel and alimented with a voltage source. As the crack propagated, the lines were cut at successive times detected with an oscilloscope (time accuracy 0.1 μs) and allowed to record the instantaneous macroscopic crack velocity $v$, with better than 10% accuracy. The variations of the quasi-static stress intensity factor $K$ were computed using 2D finite element calculations (software CASTEM 2007) on the exact experimental geometry, assuming plane stress conditions, and a constant wedge position throughout failure of the specimen. Values for the fracture energy $\Gamma$ were directly obtained from the equation of motion $\Gamma \approx (1 - v/c_R)\frac{K^2}{E}$ by combining the $v$ measurements and the $K$ calculations.

Post mortem topography images were obtained with an optical profilometer (M3D, Fogale Nanotech, ×5 objective yielding square pixels of size 1.86 μm) at various locations along the fracture surfaces in different broken specimens - each zone of observation is characterized by a given value of $K$. For each location, nine neighbouring images were gathered to provide an observation field of at least $2 \times 2$ mm$^2$, large enough to carry out statistical analyses. The presence of a highly reflective area at the focus of each conic-like marking results from plastic deformations at microcrack nucleation and allows locating unambiguously all nucleation centers (see Fig. 2A). For many microcracks, fragmentation lines focusing on the nucleation center were also observed, and helped increasing the accuracy of the location. For each marking, we made an initial guess about which microcrack triggered its nucleation. The apex of the marking was defined as the intersection between the segment linking the triggering and triggered centers and the conic-like marking. A new guess was made if the simulated marking did not resemble the observed one.

Simulation

The macroscopic crack front was initially straight, vertical and located on the left of the image. It started propagating towards the right at constant velocity (1 pixel/time step). When the macroscopic crack front reached a distance $d_n$ from the closest nucleation center, a microcrack was nucleated and made grow radially at the same velocity. The total front was then made of both the initially straight translating front and the newly created radially growing circular front. When these two coincided, propagation was continued in the unbroken part of the specimen only. Intersection points defined the conic-like marking. The same procedure was applied each time the shortest distance between the total front and another nucleation center was found to have decreased down to the distance at nucleation $d_n$ associated with this center. Edge effects were minimized in the evaluation of $A$ by considering only the times after all points of the initial front coalesced with a nucleated microcrack, and before the first point of the total front reached the right edge of the image.

We thank A. Prevost for his help with the profilometry measurements. We thank K. Ravi-Chandar for helpful discussions. We thank L. Barbier, G. Debrégeas, A. Malthe-Sorenssen, J. Mathiesen, K.J. Måloy, P. Meakin, and C. Rountree for careful reading of the manuscript. We acknowledge funding from French ANR through Grant No. ANR-05-JCJC-0088, from Triangle de la Physique through Grant No. 2007-46, from Mexican CONACYT through Grant No. 190091 and from the European Union through Marie Curie grant No. PIEF-GA-2009-237089. This Letter was supported by a Center of Excellence grant to PGP from the Norwegian Research Council.

REFERENCES

[1] Freund LB (1990) *Dynamic Fracture Mechanics* (Cambridge University Press, Cambridge, England).
[2] Fineberg J, Marder M (1999) Instability in dynamic dracture. *Phys. Rep.* 313:1-108.
[3] Ravi-Chandar K (2004) *Dynamic Fracture* (Elsevier, Amsterdam, The Netherlands).
[4] Cox B, Gao HJ, Gross D, Rittel D (2005) Modern topics and challenges in dynamic fracture. *J. Mech. Phys. Solids* 53:565-596.
[5] Livne A, Bouchbinder E, Svetlizky I, Fineberg J (2010) The near-tip fields of fast cracks. *Science* 327:1359-1363.
[6] Scheibert J, Guerra C, Célarié F, Dalmas D, Bonamy D (2010) Brittle-quasibrittle transition in dynamic fracture: An energetic signature. *Phys. Rev. Lett.* 104:045501.
[7] Ravi-Chandar K, Yang B (1997) On the role of microcracks in the dynamic fracture of brittle materials. *J. Mech. Phys. Solids* 45:535-563.
[8] Ravi-Chandar K (1998) Dynamic fracture of nominally brittle materials. *Int. J. Fracture* 90:83-102.
[9] Hull D (1999) *Fractography: Observing, measuring and interpreting fracture surface topography* (Cambridge University Press, Cambridge, England).
[10] Sharon E, Gross SP, Fineberg J (1995) Local crack branching as a mechanism for instability in dynamic fracture. *Phys. Rev. Lett.* 74:5096-5099.
[11] Sharon E, Fineberg J (1999) Confirming the continuum theory of dynamic brittle fracture for fast cracks. *Nature* 397:333-335.
[12] Bergkvist H (1974) Some experiments on crack motion and arrest in polymethylmethacrylate. *Eng. Fract. Mech.* 6:621-622.
[13] Goldman T, Livne A, Fineberg J (2010) Acquisition of inertia by a moving crack. *Phys. Rev. Lett.* 104:114301.
[14] Washabaugh PD, Knauss W (1994) A Reconciliation of dynamic crack velocity and Rayleigh-wave speed in isotropic brittle solids. *Int. J. Fracture* 65:97-114.
[15] Boudet JF, Ciliberto S, Steinberg V (1996) Dynamics of crack propagation in brittle materials. *J. Phys. II* 6:1493-1516.
[16] Rosakis AJ, Samudrala O, Coker D (1999) Cracks faster than the shear wave speed. *Science* 284:1337-1340.
[17] Kermode JR, et al. (2008) Low-speed fracture instabilities in a brittle crystal. *Nature* 455:1224-1227.
[18] Smekal A (1953) Zum Bruchvorgang bei sprodem Stoffverhalten unter ein- und mehrachsigen Beanspruchungen. *Oesterr. Ing. Arch.* 7:49.
[19] Rabinovitch A, Belizovsky G, Bahat D (2000) Origin of mist and hackle patterns in brittle fracture. *Phys. Rev. B* 61:14968.
[20] Mukhopadhyay P, Loeck M, Gottstein G (2007) A cellular operator model for the simulation of static recrystallization. *Acta Mater.* 55:551-564.
[21] Jettestuen E, Nermoen A, Hestmark G, Timdal E, Mathiesen J (2010) Competition on the rocks: Community growth and tessellation. *PLoS ONE* 5:e12820.
[22] Vestgarden JI, Shantsev DV, Galperin YM, Johansen TH (2008) Flux distribution in superconducting films with holes. *Phys. Rev B* 77:014521.
[23] Estevez R, Tijssens MGA, Van der Giessen E (2000) Modeling of the competition between shear yielding and crazing in glassy polymers. *J. Mech. Phys. Solids* 48:2585-2617.
[24] Persson BNJ, Brener EA (2005) Crack propagation in viscoelastic solids. *Phys. Rev. E* 71:036123.
[25] Buehler MJ, Abraham FF, Gao, HJ (2003) Hyperelasticity governs dynamic fracture at a critical length scale. *Nature* 426:141-146.

[26] Rountree CL, et al. (2002) Atomistic aspects of crack propagation in brittle materials: Multimillion atom molecular dynamics simulations. *Ann. Rev. Mater. Res.* 32:377-400.
[27] Moore DE, Lockner DA (1995) The role of microcracking in shear-fracture propagation in granite. *J. Struct. Geol.* 17:95-114.
[28] Kobchenko M, Panahi H, Renard F, Dysthe DK, Malthe-Sorenssen A, Mazzini A, Scheibert J, Jamtveit B, Meakin P (2011) 4D imaging of fracturing in organic-rich shales during heating. *J. Geophys. Res.* doi:10.1029/2011JB008565, in press.
[29] Nalla RK, Kinney JH, Ritchie RO (2003) Mechanistic fracture criteria for the failure of human cortical bone. *Nat. Mater.* 2:164-168.

# Supporting Information

**SI Text**

Relative speed of interacting microcracks as deduced from the geometry of conic-like markings.

Figure S2(A) depicts the interaction between two microcracks growing radially at velocities $c_1$ and $c_2$, respectively. Calling $\Delta$ the distance between the two nucleation centers and $\tau$ the time interval between the nucleations of the first and second microcracks, the equation describing the successive locations of intersection points (and hence the conic-like marking) is given by:

$$\frac{y}{\Delta} = \pm\sqrt{-\left(\frac{x}{\Delta}\right)^2 + \frac{c^2+1}{(c^2-1)^2}\left(c^2\left(\frac{c_1\tau}{\Delta}\right)^2 - \frac{2x}{\Delta}\right) + \frac{1}{(c^2-1)^2} + \frac{2c^2}{(c^2-1)^2}\frac{c_1\tau}{\Delta}\sqrt{\left(1-\frac{2x}{\Delta}\right)(c^2-1) + c^2\left(\frac{c_1\tau}{\Delta}\right)^2}}, \quad (S1)$$

where $c=c_2/c_1$ and $(x,y)$ are expressed in a frame the origin of which is the first nucleation center and the $x$-axis goes through both centers (see Fig. S2(A)). As can be seen in Figs. S2(B) and (C), qualitatively different shapes are predicted by this equation depending on the velocity ratio $c_2/c_1$: True (mathematical) conics when $c_2/c_1=1$, egg shapes when $c_2/c_1<1$ and flared shapes when $c_2/c_1>1$. The fact that the markings observed on the post-mortem fracture surfaces look like conics hence suggests a velocity ratio close to unity.

To assess quantitatively the value of the velocity ratio, we directly extracted $c_2/c_1$ for each pair of interacting microcracks from the marking's geometry on the fractographic images (see Fig. 1(B) and Fig. S4(top)). Note that seemingly continuous conic-like markings often result from the successive interactions between more than two microcracks, and are hence actually made of several branches, each of them associated with a single pair of interacting microcracks. The analysis procedure we developed is the following: (i) the nucleation centers of two interacting microcracks are selected on the image; (ii) the apex of the associated marking is determined as the intersection of the segment relating the two centers and the marking's branch lying in between; (iii) the marking's shape predicted by equation (S1) is plotted for various ratio $c_2/c_1$ while adjusting $c_1\tau/\Delta$ so that the apex position remains fixed; and (iv) the value $c_2/c_1$ that best fits the experimental marking is selected. Typical examples of the cumulative distributions obtained for $c_2/c_1$ are presented in Fig. 2(B) and Fig. S4(bottom). For all the fractographic images analysed, the distributions were found to be roughly Gaussian, with mean values around 0.98-0.99 and standard deviation of 0.02-0.03, irrespective of the density $\rho$ of conic-like markings. Note that local stationarity of the average crack front over the millimeter length-scales of the analyzed fractographic images should imply a geometric mean value of $c_2/c_1$ strictly equal to 1. The dispersion smallness around 1 allows identification of this geometric mean with the standard arithmetic one. The observation of a mean value systematically slightly smaller than unity is attributed to an initial accelerating transient in microfracturing events that is not taken into account to derive Eq. S1.

It is then justified to assume that all microcracks propagate at the same constant velocity $c_1=c_2=c_m$ on a given fractographic image. Equation (S1) can then be simplified to:

$$\frac{y}{\Delta} = \pm\sqrt{4\frac{d_n(2\Delta-d_n)}{(\Delta-2d_n)^2}\left(\frac{x}{\Delta}\right)^2 - 4\frac{d_n(2\Delta-d_n)}{(\Delta-2d_n)^2}\left(\frac{x}{\Delta}\right) + \frac{d_n(2\Delta-d_n)}{(\Delta-2d_n)^2} - 4\left(\frac{d_n}{\Delta}\right) + 4\left(\frac{d_n}{\Delta}\right)^2}, \quad (S2)$$

where $d_n = \Delta - c_m\tau$ is the distance between the triggering front and the nucleation point at the instant of microcrack nucleation and is equal to twice the distance between the marking's apex and focus. Note that, contrary to Eq. (S1), Eq. (S2) describes a true mathematical conic (see Fig. 1(C)), the eccentricity and focal parameter of which depend on both $d_n$ and $\Delta$.

Variation of microcrack density with stress intensity factor

The $K$ dependency of $\rho$ (Fig. 2(A)bottom) has been understood in Ref. *(S1)* by assuming the material to contain a population of local weak zones, so-called "source-sinks" (SS). Each one is able to turn into a microcrack provided two conditions are met: (i) the local stress at the considered SS reaches a threshold value $\sigma_*$ (smaller than the local yield stress $\sigma_Y$) and (ii) the SS is located at a distance from the crack front larger than $d_a$. Calling $\rho_v$ the SS volume density, the surface density $\rho$ is then equal to the number of activated SS beyond $d_a$ per unit of fracture area, *i.e.* $\rho_v \{h_\perp - 2d_a - \rho V\}$, where $h_\perp$ is the thickness (size orthogonal to the fracture plane) of the FPZ, *i.e.* the layer in which the stress is larger than $\sigma_*$. $V$ is the excluded volume around nucleated microcracks. The universal square root singular form taken by the elastic stress field around the tip of the growing crack gives $h_\perp = K_d^{\,2} / \alpha_2 \sigma_*^2$ where $\alpha_1$ is a dimensionless constant close to unity, and $K_d$ relates to $K$ via $K_d = k(c_m) K$ with $k(c_m) = (1 - c_m / c_R) / \sqrt{1 - c_m / c_D} \approx 0.81$ (Ref. *(S3)*). Finally, one gets:

$$\rho = C\left(K^2 - K_a^2\right) \text{ with } C = \frac{\rho_v}{1 + \rho_v V} \frac{k^2(c_m)}{\alpha_1 \sigma_*^2} \text{ and } K_a^2 = \frac{2\alpha_1 \sigma_*^2 d_a}{k^2(c_m)} \tag{S3}$$

The parameter $K_a$ represents then the value of $K$ at the onset of microcracking. Below this value, microcracks cannot nucleate and $\rho=0$. Equation (S3) is found to reproduce the experimental data fairly well (see Fig. 1(A)bottom in the main text) up to a value $K_b \approx 3.1 \text{MPa.m}^{1/2}$ which is associated with the microbranching onset (see Ref. *(S1)*). The loss of agreement above $K_b$ is actually expected since LEFM is invalid in the presence of microbranches (see Ref. *(S2)*). The fitted parameters $C$ and $K_a$ are found to be $C = 9.0 \pm 0.5 \times 10^6 \text{MPa}^{-2}.\text{m}^{-3}$ and $K_a = 2.1 \pm 0.1 \text{MPa.m}^{1/2}$, respectively.

Variation of mean distance at nucleation with microcrack density

In the scenario invoked above, the relation between the mean distance at nucleation $\overline{d_n}$ and $K_d$ is deduced from the universal square root singular form taken by the elastic stress field around the tip of the growing crack (see *e.g.* Ref. *(S3)*). The mean distance $\overline{d_n}$ at which the stress level reaches $\sigma_*$ then reads $\overline{d_n} = K_d^{\,2} / \alpha_2 \sigma_*^2$ where $\alpha_2$ is a dimensionless constant close to unity. This relation together with Eq. (S3) yields :

$$\overline{d_n} = \frac{\alpha_1}{\alpha_2} \frac{1 + \rho_v V}{\rho_v} \rho + 2 \frac{\alpha_1}{\alpha_2} d_a \tag{S4}$$

This equation is found to reproduce the experimental data fairly well (see Fig. 2(C)bottom) up to the point where $\rho$ reaches the value $\rho_{sat} = 45.5 \text{mm}^{-2}$ defined in Fig. 2(A)bottom and identified with the microbranching onset. The fitted parameters are found to be $\alpha_1 (1 + \rho_v V) / \alpha_2 \rho_v \approx 10^6 \mu\text{m}^3$ and $2\alpha_1 d_a / \alpha_2 \approx 4.5 \mu\text{m}$.

Saturation of $\overline{d_n}$ and avalanches

Figure S6 superimposes the variation of the mean distance at nucleation $\overline{d_n}$ with $\rho$ (presented in Fig. 2(C)bottom) to the variation of the mean nearest-neighbour distance $<\Delta r>$ with $\rho$. For a Poissonian distribution, $<\Delta r> = 1/(2\sqrt{\rho})$ while the variance is given by $\sigma^2_{\Delta r} = (4-\pi)/(4\pi\rho) \approx 0.068/\rho$. When $\rho$ becomes of the order of $\rho_{sat}$, the distance between some nucleation centers and the centers of their triggering microcrack becomes smaller than the distance necessary for them to nucleate. Hence, both centers will open almost simultaneously (within the same time step), and an avalanche will occur. This steric effect yields an effective mean distance at nucleation $\overline{d_n}$ (as measured from fracture surfaces) that saturates above $\rho_{sat}$ at a value $\overline{d_n} \approx 50 \mu m$. The proportion of micro-cracks involved in nucleation avalanches increases significantly in the vicinity of $\rho_{sat}$ (solid squares in Fig. S6). However, this proportion remains small, less than 7% over the explored range of densities. This explains why the mean-field model yielding Eq. (1) captures so well the increase of the acceleration factor with $\rho$, even above $\rho_{sat}$ (see Fig. 4(B)bottom).

Edge effects in the deterministic reconstruction

Once the growth rule $c_2=c_1$ has been ascertained experimentally, and once the nucleation center location and distance at nucleation of all individual microcracks have been determined from image analysis, the main source of mismatch between the reconstruction and the actual fracture surfaces are errors in the time succession of nucleation events. Such errors are unavoidable when using partial images (a few mm$^2$) of the complete fracture surface (a few cm$^2$). Along the top and bottom sides of the image, we lack for the possibility of being triggered by microcracks outside the field of view. From the left side of the image, we lack for a realistic initial front shape which would contain all the information about the precise instants at which the leftmost centers have to be nucleated during the reconstruction. Since no information could be obtained about this initial front shape, we arbitrarily chose a straight vertical front as an initial condition.

In Fig. S7, we illustrate the degree of sensitivity of the reconstruction results to changes in the initial front shape, by running the simulation with a sinusoidal initial front. Its period is chosen to roughly match the average vertical distance between simultaneously propagating microcracks. Its amplitude is set to the mean value of the standard deviation of the horizontal location of the total front as obtained using an initial straight front. The same sinusoidal shape, but translated vertically by half a period, was also tested. As the front involves more generations of microcracks, the patterns become more alike, and hence less affected by the initial conditions. However, some differences can propagate over the whole image. We emphasize that nearly perfect reconstructions could have been obtained through the analysis of the images which include the region where the first microcracks nucleate, because it would have allowed for the determination of the actual initial front shape.

The crucial point for our study is to check that the reconstruction errors induced by edge effects do not quantitatively affect the value $A$ of the acceleration factor. Fig. S7 shows that, for the three very different initial front shapes tested, $A$ is found constant within less than 0.6%. This means that the average velocity measurement hardly depends on edge effects.

These results were found to be robust to changes in period and amplitude of the sinusoidal initial front.

Lattice model

The variations of the ratio between the macro and micro-scale velocities, *i.e.* the acceleration factor $A$, as a function of microcrack density $\rho$ can be captured by a simple mean-field model. In the model, the nucleation centers are placed at the nodes of a square lattice, so that the distance $\ell$ between two neighbouring centers is kept constant and equal to $\ell = 1/\sqrt{\rho}$. This model arrangement is depicted in Fig. S8, where $x$ and $z$ axes are chosen parallel to mean crack propagation direction and to mean crack front, respectively. At $t = 0$, the leftmost nucleation centers are turned into microcracks and start to grow radially at velocity $c_m$. The fronts then trigger the nucleation of the next centers when the shortest distance between the nucleation centers and the fronts reaches a distance of $\overline{d_n}$. The nucleated microcracks then grow radially, coalescing with each other and with the primary crack. The new microcracks trigger the nucleation of the next microcracks, and so on. Because of the invariance to translation along the $z$-axis the crack can be considered to propagate in the $x$ direction only (see Fig. S8). When the main front has travelled over a distance $L = Ac_m t$ along this line, it has triggered $L/\ell = L\sqrt{\rho}$ micro-cracks. And because each coalescence with a microcrack makes the rightmost point jump over a distance $\overline{d_n}$, while the crack velocity is $c_m$ between these coalescence events, one also gets $L = c_m t + L\overline{d_n}\sqrt{\rho}$. From the two expressions for $L$, it can be deduced that $A = 1/(1 - \overline{d_n}\sqrt{\rho})$.

In real materials, the centers are not aligned along lines parallel to the direction of mean crack propagation but are distributed randomly. We thus propose to modify this equation into:

$$A = \frac{1}{1 - b\overline{d_n}\sqrt{\rho}} \tag{S5}$$

where the geometrical constant $b$ (expected to be close to 1) accounts for the projection onto the $x$ axis of (i) the real distance between successive nucleation centers and (ii) the distance jumped during a coalescence event with a non-aligned micro-crack. This equation is the one proposed in the main text (Eq. (1)).

**Fig. S1: Sketch of the experimental setup.**

Figure S1: Sketch of the so-called wedge splitting geometry used to grow dynamic cracks in PMMA (see Materials and Methods and Ref. (S1)).

**Fig. S2: Geometry of conic markings.**

Figure S2: (A) Sketch underlying equation (S1). A first microcrack nucleates at time $t=0$ at frame origin $O_1$, and subsequently grows at velocity $c_1$. A second microcrack nucleates at time $\tau$ at point $O_2$ of coordinates ($x=\Delta$, $y=0$), and subsequently grows at velocity $c_2$. The two grey circles correspond to both microcrack fronts at time $t$, if microcrack interaction was ignored. In reality, the intersection of the two fronts leaves on the fracture surface a marking that develops as the blue curve as time $t$ increases. In dimensionless coordinates $x/\Delta$ and $y/\Delta$, the marking aspect is set by the ratios $c_2/c_1$ and $c_1\tau/\Delta$. The forms obtained for $c_2/c_1=0.9$ (blue), $c_2/c_1=1$ (red) and $c_2/c_1=1.1$ (green) are plotted in (B) and (C), for $c_1\tau/\Delta=0.05$ and $c_1\tau/\Delta=0.2$ respectively.

**Fig. S3: Spatial distribution of nucleation centers.**

Figure S3: Cumulative distribution of the number of nucleation centers contained in square regions of lateral size $a$, for each of the eight fractographic images analyzed. The cumulative distribution is defined as the proportion of square regions containing a number of nucleation centers which is strictly superior to the value in abscissa. Four values of $a$ were chosen, namely $a = 100, 200, 300$ and $400$ $\mu$m. In each graph, solid lines represent Poisson fits $P(n) = \sum_{k=0}^{n}(\rho a^2)^k / k!$ where the fitting parameter $\rho$ is the same in all four curves and hence defines the surface density of centers in each image. Note that the same Poissonnian distribution was assumed in Ref. *(S4)*. The stress intensity factor $K$ applying on the macroscopic crack front at these points was computed using finite element calculations (Materials and Methods). Its value together with the fitted value $\rho$ is reported in each graph inset.

**Fig. S4: Distribution of microcrack relative velocity**

Figure S4: Direct extraction of the relative speed between two interacting microcracks. Top: Typical examples of investigated zones (985×745$\mu$m$^2$ in size) at three different microcrack densities. Each conic branch has been attributed a given color and the nuclei of the two corresponding interacting microcracks has been joined by a dotted segment of the same color. Note that a conic mark is often made of several of these branches. The ratio $c_2/c_1$ is the only adjustable parameter in equation (S1) to determine the branch geometry once the nuclei position and the branch apex are set. Bottom: Corresponding distributions for $c_2/c_1$. In the three cases, the distributions are found to fit normal distributions of mean value ~0.98-0.99 and standard deviation ~0.03-0.04, irrespective of $\rho$.

**Fig. S5: Distribution of distances at nucleation.**

Figure S5: Cumulative distribution, $P$, of the distance $d_n$ between the triggering crack and the nucleation center at the time of nucleation, determined from each of the analyzed fractographic images. In each graph, the solid line shows the fit of the form $P(d_n) = ((d_{max} - d_n) / (d_{max} - d_{min}))^2$, where $d_{min}$ and $d_{max}$ are positive quantities. The fitted values $d_{min}$ and $d_{max}$ together with the surface density of nucleation centers $\rho$ are reported in inset in each graph. $d_{min}$ decreases with $\rho$ and becomes equal to zero when $\rho$ is larger than 36 mm$^{-2}$. $d_{max}$ increases with $\rho$ over the whole explored range. Neglecting the small value of $d_{min}$ allows us to use a single parameter (*e.g.* $\overline{d_n}(\rho)$) to define the whole distribution and its variations.

**Fig. S6: Saturation of $\overline{d_n}$ and avalanches.**

Figure S6: Evolution of the fitted mean distance at nucleation, $\overline{d_n}$, (thick red line, see Fig. 2(C)bottom), as a function of $\rho$. It is compared to the mean nearest-neighbor distance in a Poissonian distribution $1/(2\sqrt{\rho})$ (black solid line, error bars correspond to ± one standard deviation $\approx 0.26/\sqrt{\rho}$). Solid squares indicate the proportion of micro-cracks involved in avalanches, as computed from the reconstruction.

**Fig. S7: Influence of edge effects on the reconstruction and average velocity**

Figure S7: Top line: Fracture surface images (grey level) compared to the reconstructed conic marks (red lines), for $\rho$=45.0mm$^{-2}$ ($K \simeq 3.65$MPa.m$^{1/2}$). Three different initial conditions were used. A: Straight vertical line. B: Vertical sinusoidal shape with a period of 186μm and a peak-to-peak amplitude of 242μm. C: Same sinusoidal shape, but translated vertically over half a period. In all cases, red dots indicate nucleation centers. Bottom: Time evolution (scaled by $c_m$) of the average location of the simulated crack front for the three different initial front shapes. For each curve, the dashed line is a linear fit of the data between the two black dots. The fitted slopes between the black dots, which directly give the value of the acceleration factor $A$, are 1.679±0.001 (straight), 1.666±0.001 (sinus 1) and 1.687±0.001 (sinus 2). This shows that the value of $A$ is hardly sensitive to edge effects in the reconstruction.

**Fig. S8: Sketch of the lattice model.**

Figure S8: Sketch of the mean field model used to establish the relationship between the acceleration factor $A$, the microcrack density $\rho$, the mean distance at nucleation $\overline{d_n}$ and the microscopic velocity $c_m$ of micro-crack growth (Eq. (1) in the main text). The nucleation centers are placed on the nodes of a square lattice. The crack front is plotted at 10 successive times separated by a constant interval $\overline{d_n}/c_m$ (from blue to red). At each time, the rightmost point of the front is projected along the $x$ axis (thick black ticks on the $x$ axis). It propagates at a constant velocity $c_m$ between coalescence events, and jumps over a distance $\overline{d_n}$ at coalescence.

**Movie S1: Reconstructed fracture dynamics for $\rho = 27.5$ mm$^{-2}$**

Movie S1: Movie of the reconstructed microscopic dynamics of crack propagation and micro-cracking events, for a density of micro-cracks $\rho = 27.5$ mm$^{-2}$. It shows a region of size $\simeq 2500\mu$m $\times$ $2400\mu$m during $\simeq 9.8$ $\mu$s.

**Movie S2: Reconstructed fracture dynamics for $\rho = 64.6$ mm$^{-2}$**

Movie S2: Movie of the reconstructed microscopic dynamics of crack propagation and micro-cracking events, for a density of micro-cracks $\rho = 64.6$ mm$^{-2}$. It shows a region of size $\simeq 3400$ $\mu$m $\times 2500$ $\mu$m during $\simeq 8.5$ $\mu$s.

**References**

S1. Scheibert J, Guerra C, Célarié F, Dalmas D, Bonamy D (2010) Brittle-quasibrittle transition in dynamic fracture: An energetic signature. Phys. Rev. Lett. 104:045501.
S2. Sharon E, Fineberg J (1999) Confirming the continuum theory of dynamic brittle fracture for fast cracks. Nature 397:333-335.
S3. Freund LB (1990) Dynamic Fracture Mechanics (Cambridge University Press, Cambridge, England).
S4. Ravi-Chandar K, Yang B (1997) On the role of microcracks in the dynamic fracture of brittle materials. J. Mech. Phys. Solids 45:535-563.

NOTE: The following pages correspond to figures S1 to S8, in the order.

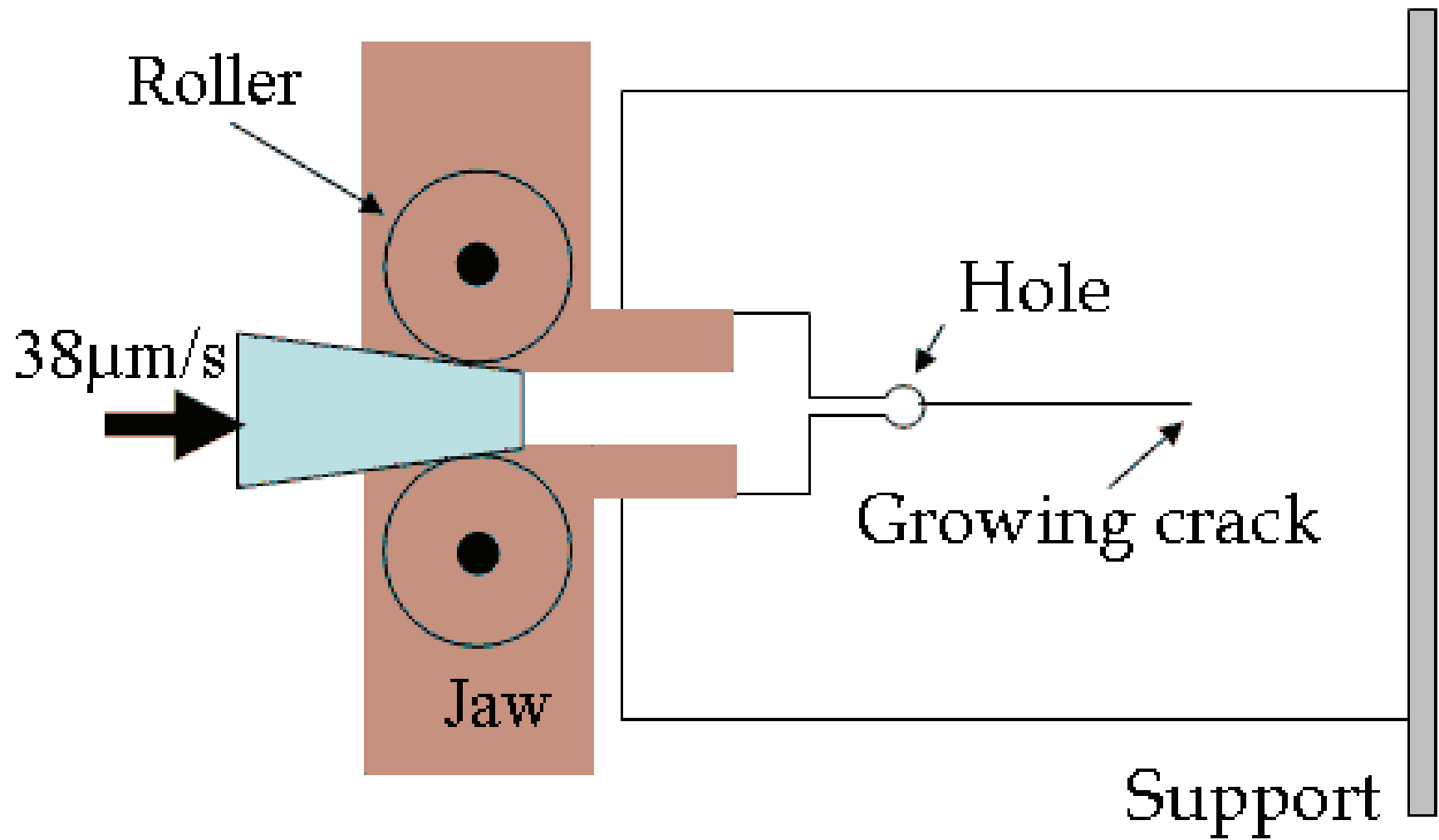

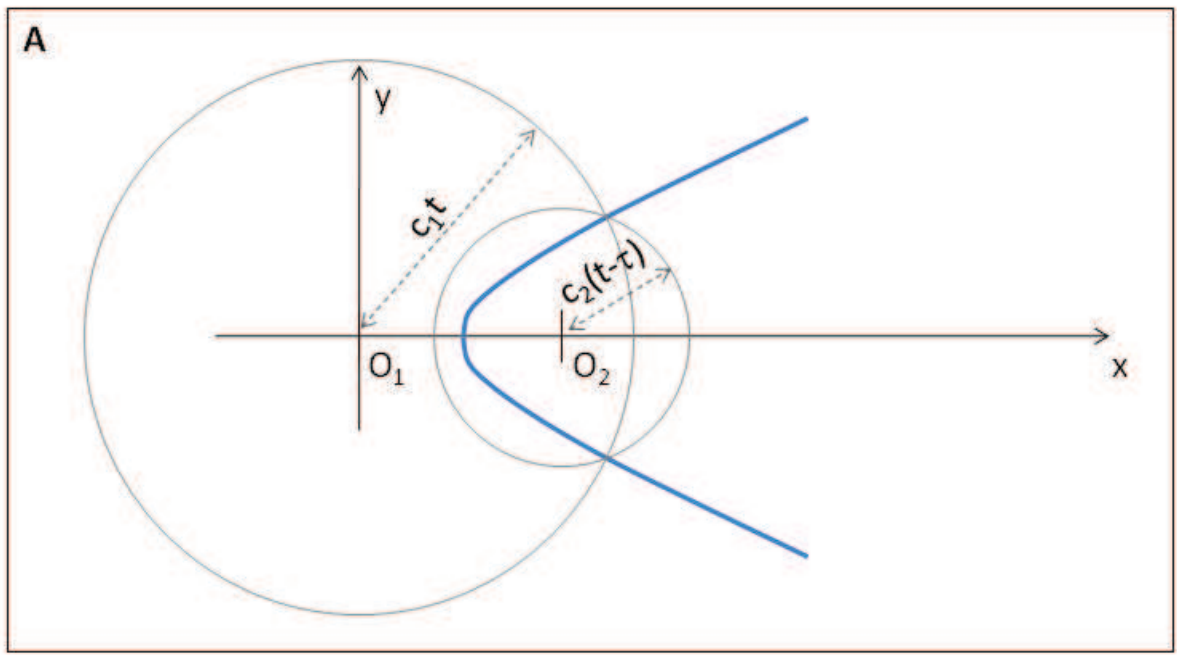
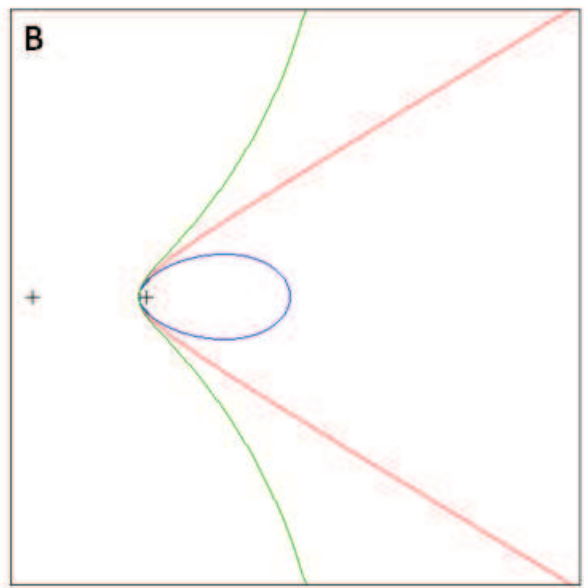
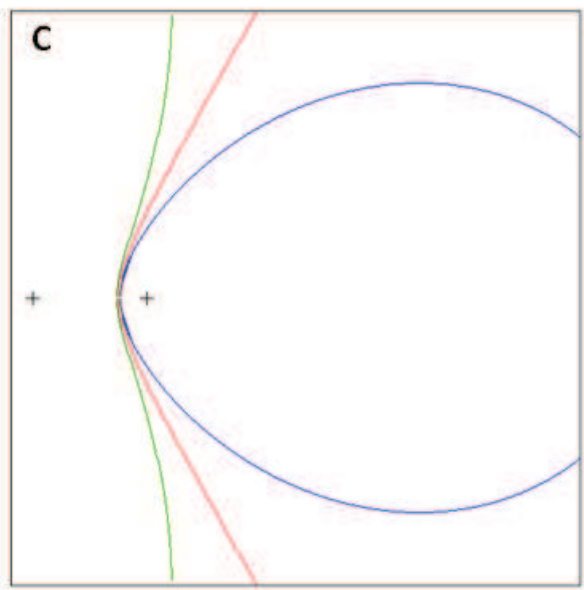

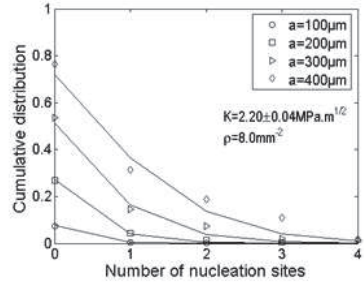
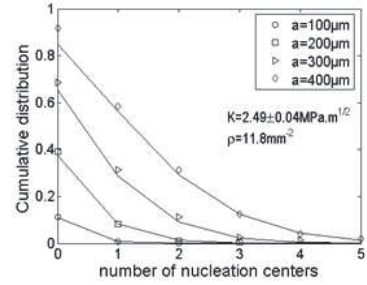
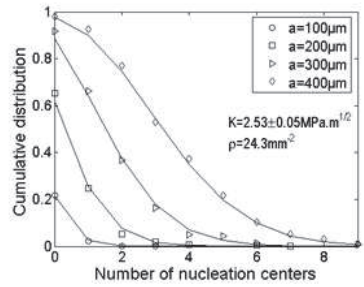
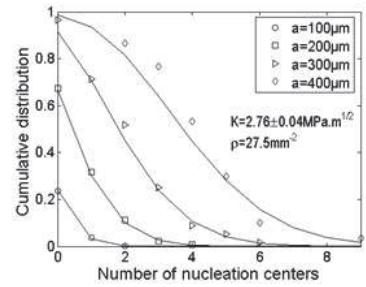
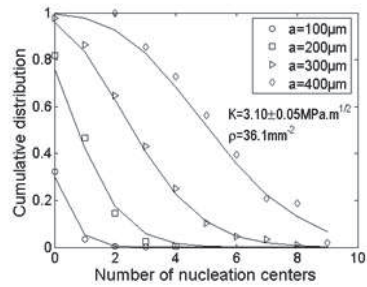
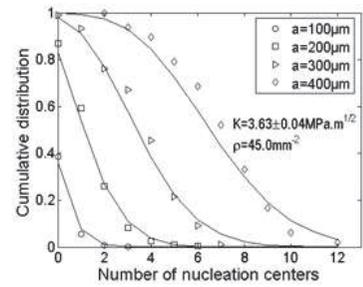
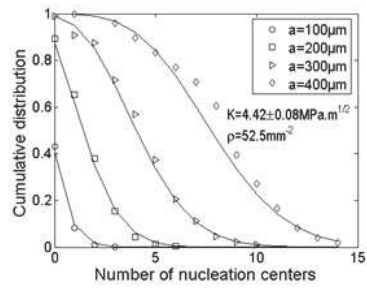
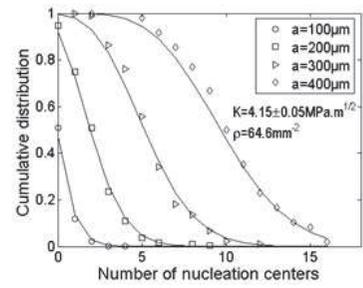

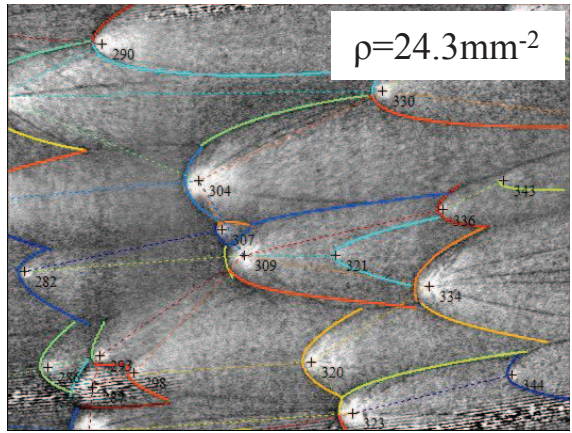 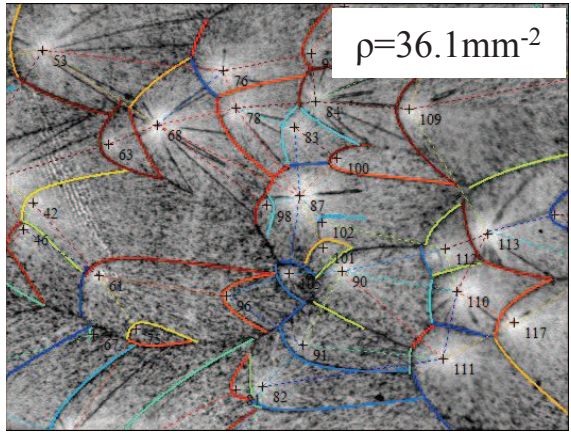 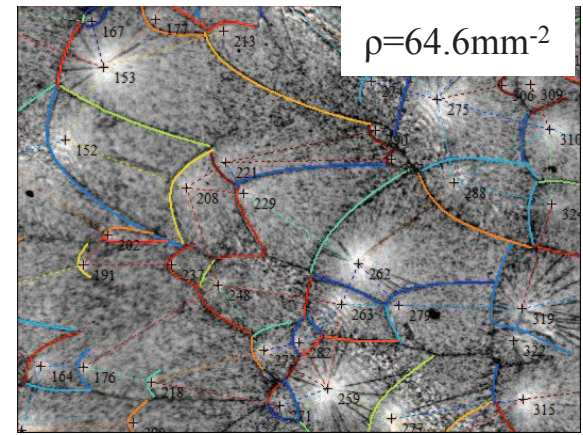

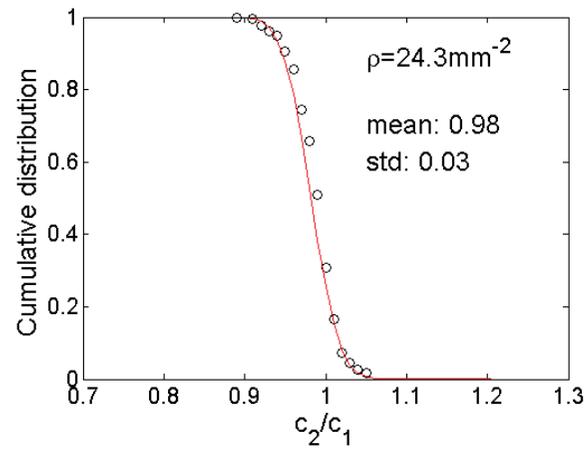 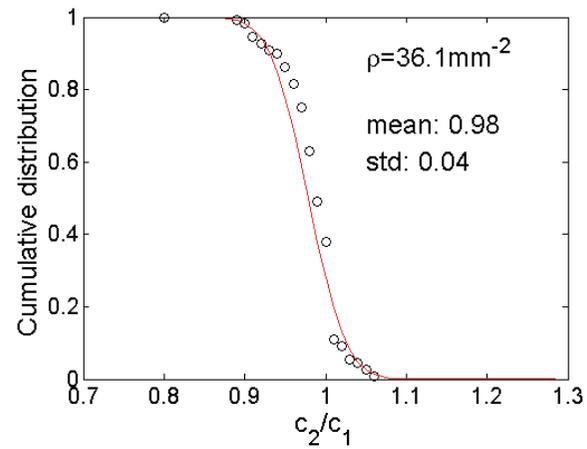 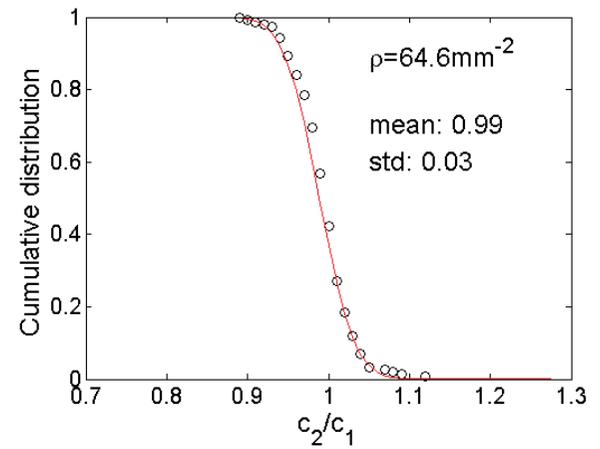

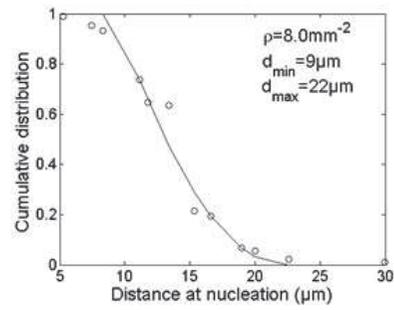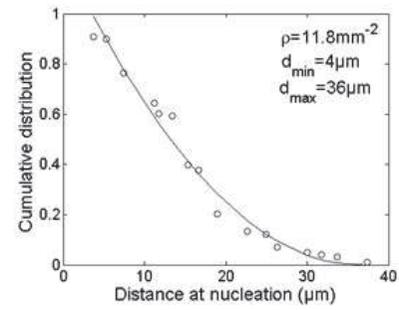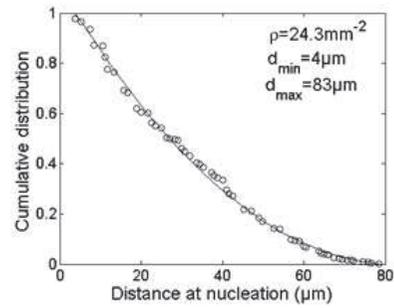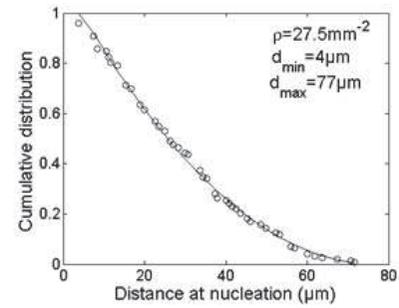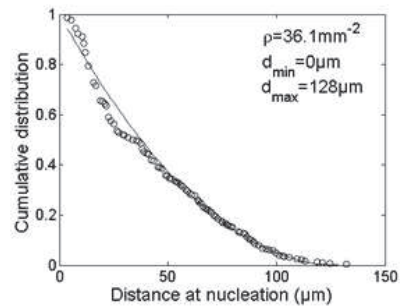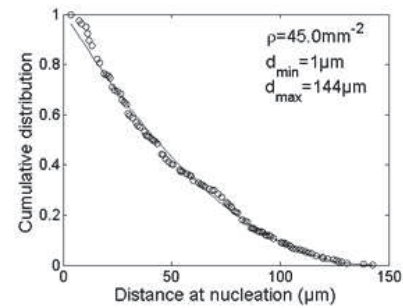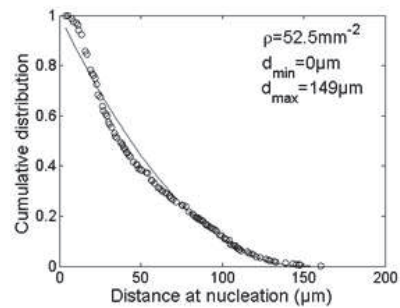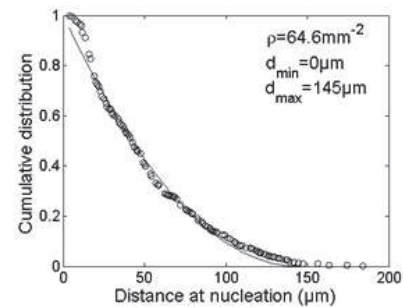

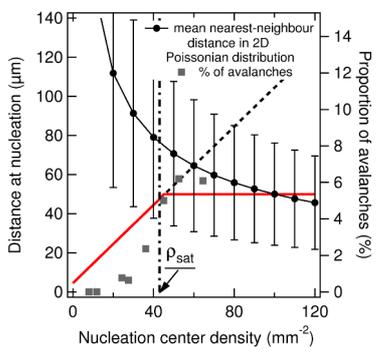

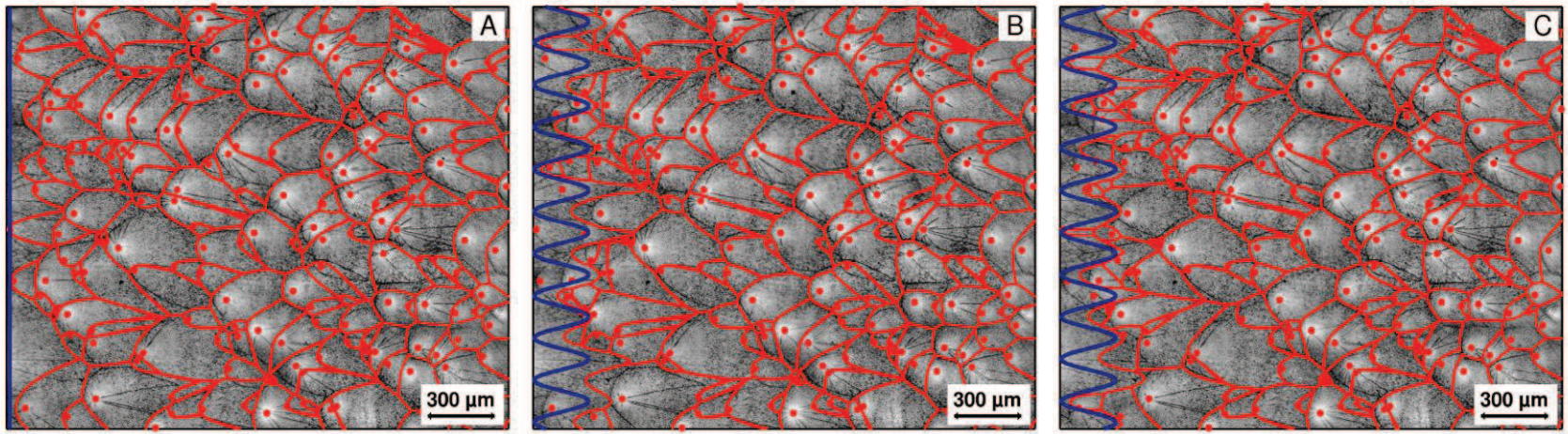
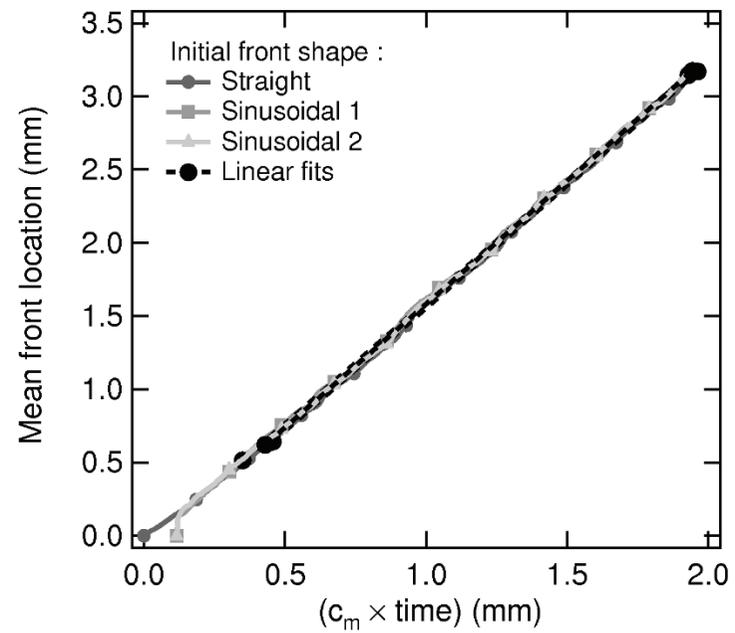

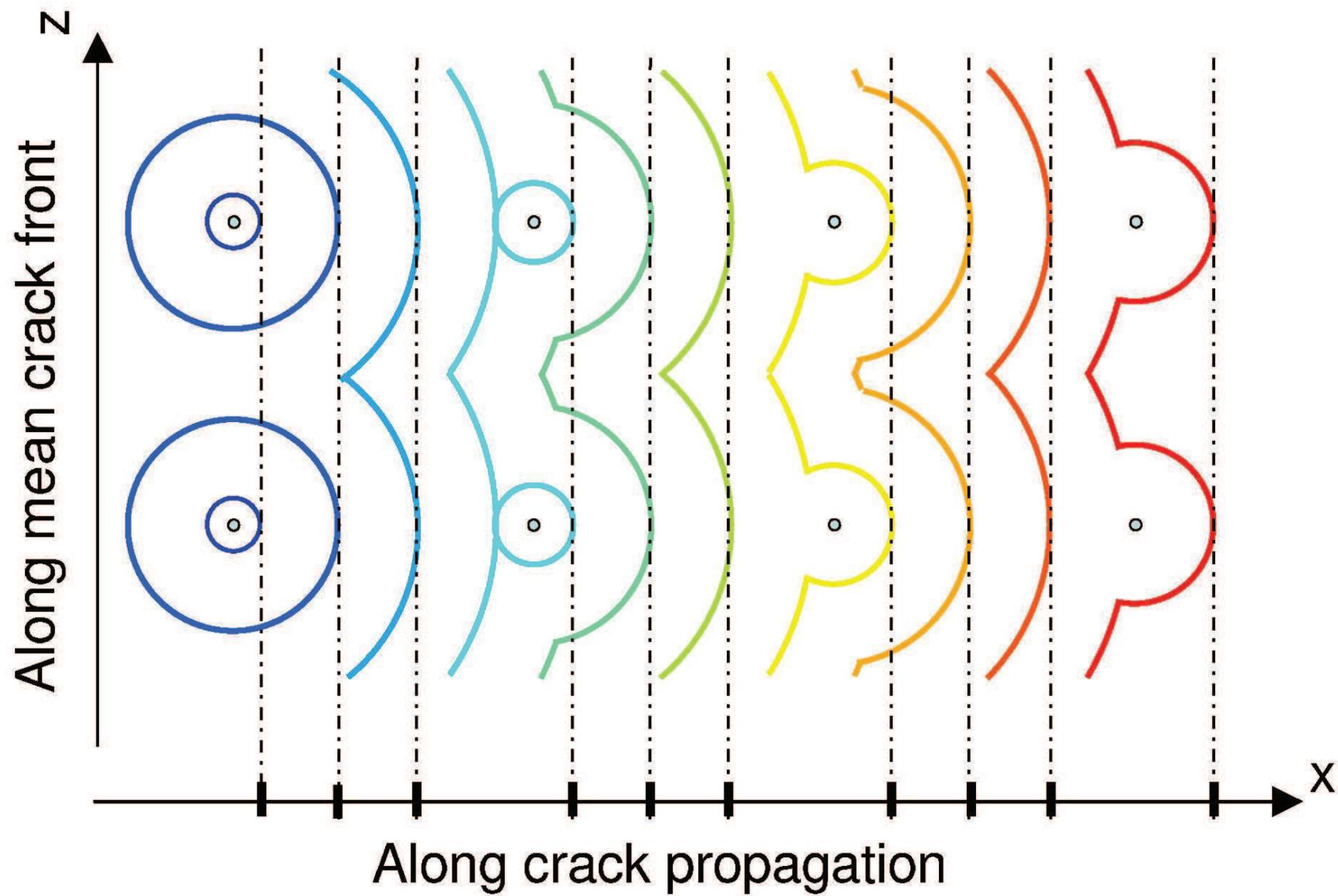